\documentclass[aps,twocolumn,showpacs,floatfix,superscriptaddress]{revtex4}
\usepackage{graphicx}
\usepackage{amsmath}
\usepackage{amsfonts}
\usepackage{amssymb}

\begin{document}
\def\be{\begin{equation}}
\def\ee{\end{equation}}
\def\bea{\begin{eqnarray}}
\def\eea{\end{eqnarray}}
\def\bma{\begin{mathletters}}
\def\ema{\end{mathletters}}
\newcommand{\one}{\mbox{$1 \hspace{-1.0mm}  {\bf l}$}}
\newcommand{\eins}{\mbox{$1 \hspace{-1.0mm}  {\bf l}$}}
\def\C{\hbox{$\mit I$\kern-.7em$\mit C$}}
\newcommand{\tr}{{\rm tr}}
\newcommand{\half}{\mbox{$\textstyle \frac{1}{2}$}}
\newcommand{\shalf}{\mbox{$\textstyle \frac{1}{\sqrt{2}}$}}
\newcommand{\ket}[1]{ | \, #1  \rangle}
\newcommand{\bra}[1]{ \langle #1 \,  |}
\newcommand{\proj}[1]{\ket{#1}\bra{#1}}
\newcommand{\kb}[2]{\ket{#1}\bra{#2}}
\newcommand{\bk}[2]{\langle \, #1 | \, #2 \rangle}
\def\II{I(\{p_k\},\{\rho_k\})}
\def\ss{{\cal K}}
\tolerance = 10000




\title{Fermionic atoms in optical superlattices}

\author{B. Paredes}
\affiliation{Max--Planck Institute for Quantum Optics,
Garching, Germany}
\author{C. Tejedor}
\affiliation{Dept. F\'{\i}sica Te\'orica de la Materia Condensada, 
Universidad Aut\'onoma de Madrid, Spain}
\author{J. I. Cirac}
\affiliation{Max--Planck Institute for Quantum Optics,
Garching, Germany}

\begin{abstract}

Fermionic atoms in  an optical {\em superlattice} can  realize a very peculiar
Anderson lattice model in which  impurities interact with each other through a
{\em  discretized} set of  delocalized levels.   We investigate  the interplay
between Kondo effect and magnetism  under these finite-size features.  We find
that Kondo effect  can dominate over magnetism depending on  the parity of the
number of particles per discretized  set. We show how Kondo-induced resonances
of measurable size can be observed through the atomic interference pattern. 

\end{abstract}

\date{\today}
\pacs{03.75.Fi, 03.67.-a, 42.50.-p, 73.43.-f } \maketitle

Cold atoms  in optical  lattices are  capturing a lot  of attention  from both
experimental  \cite{Bloch1,   Bloch2}  and  theoretical   sides  \cite{Dieter,
Dieter2,  Lukin,  Paredes, Luis,  Dieter3,  Jane}.   This  interest is  highly
motivated  by  the  possibility   of  investigating  the  domain  of  strongly
correlated phenomena, the interaction  effects (typically small in free space)
being enhanced due  to the periodic confinement. As a  unique feature of these
atomic systems, the full control  of the system's parameters allows to explore
several fascinating  directions.  On the  one hand, atoms in  optical lattices
can  be  used  to  provide  illuminating  and  critical  insight  into  models
describing   strongly  correlated   systems.   For   example,   quantum  phase
transitions in both bosonic  \cite {Dieter} and fermionic \cite{Lukin} Hubbard
models,  as well as  in spin  Hamiltonians \cite{Jane},  can be  explored with
unprecedent  control.   On  the other  hand,  in  what  perhaps is  even  more
challenging,  exotic scenarios  can be  created in  which  strongly correlated
phenomena  may  occur under  novel  conditions.   A  variety of  possibilities
already accessible  experimentally (different lattice  topologies created with
superpositions  of  multiple   laser  beams  \cite{superlattice},  independent
periodic  potentials  for  different  internal  atomic  states  \cite{Dieter2,
Bloch2}, interactions controlled  by Feshbach resonances \cite{Feshbach}, etc)
can be combined, promising new ways to strongly entangle atomic ensembles.

In this  letter we  study the physics  of fermionic  atoms in an  optical {\em
superlattice} \cite{superlattice}.   We will show that the  system realizes an
Anderson  Lattice  Hamiltonian  (ALH)   \cite{Anderson}.   The  ALH  has  been
extensively  studied   in  the   context  of  strongly   correlated  electrons
\cite{Fulde,Hewson},  and is  known to  capture the  physics of  a  variety of
strongly  correlated  phenomena,  from   Kondo  effect  \cite{Fulde}  to  RKKY
magnetism \cite{Fulde, Hewson}.  Typical condensed matter systems described by
Anderson  models   are  metallic  or   intermetallic  compounds  with   a  low
concentration  of magnetic  impurities.  The  usual scenario  is then  that of
impurities  located far  from  each other,  each  of them  coupled  to a  {\em
continuum} of delocalized electrons.  In an interesting volte-face, atoms in a
superlattice  naturally realize a  quite different  situation, allowing  us to
investigate  a very  peculiar regime.   For realistic  experimental situations
supersites (that will play the role of impurities) will be separated typically
by a small  number of lattice sites.  Therefore,  (if, for instance, tunneling
is  only  allowed  along one  direction),  the  system  realizes an  array  of
impurities connected through small ``islands'' with a {\em discretized} set of
levels.   The  situation  resembles  that  of  an array  of  the  Kondo  boxes
theoretically studied  in \cite{Jan}, where  impurities can now  interact with
each  other through  the intermediate  ``conducting islands''.   We  will show
that, in such a situation, both Kondo effect and the competition between Kondo
effect  and magnetism  are strongly  affected by  finite-size effects,  with a
remarkable enhancement of the Kondo temperature.  We explain how to induce and
observe  the  strongly correlated  effects  we  predict  by combining  several
different techniques.

We consider a gas of fermionic  atoms embedded in a superlattice of period $L$
with  potential depth $V_{0}$  for ``normal''  sites and  $V_{0}^{\prime}$ for
``supersites'' (see  Fig.1).  We  assume that two  kinds of atoms  are present
(generalized spin $\sigma={\uparrow, \downarrow}$).
\begin{figure}[h]
\includegraphics[height=4.33cm,width=6.5cm]{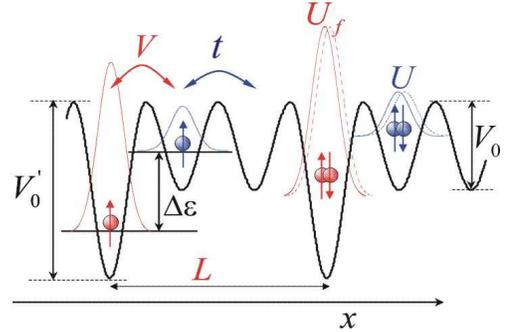}
\caption{AHM for atoms in a superlattice (see text).}
\end{figure}  
For sufficiently low temperatures atoms will be confined to the
lowest Bloch  band of the superlattice and  the system can be  described by an
ALH of the form
\begin{eqnarray}
H_{ALH}&=&
-t\hspace{-0.1cm}\sum_{<\ell,\ell^{\prime}>\sigma}\hspace{-0.1cm}
c^{\dagger}_{\ell\sigma}c^{}_{\ell^{\prime}\sigma}
+U\sum_{\ell}n_{\ell\uparrow}n_{\ell \downarrow} 
-\Delta \epsilon \sum_{s\sigma}n_{s\sigma}^{f} \nonumber \\
&& \hspace{-0.3cm}+\;\;U_{f}\sum_{s}\;n_{s\uparrow}^{f}n_{s\downarrow}^{f}
+V\hspace{-0.1cm}\sum_{<\ell,s>\sigma}\left(f^{\dagger}_{s\sigma}c^{}_{\ell
\sigma}+ \mbox{h.c} \right),
\label{ham}
\end{eqnarray}
where $c_{\ell\sigma},f_{s\sigma}$ are  fermionic operators that annihilate an
atom  with  spin state  $\sigma$  on normal  site  $\ell$  and supersite  $s$,
respectively,  and  $n_{\ell \sigma}=c^\dagger_{\ell\sigma}c^{}_{\ell\sigma}$,
$n_{s\sigma}^{f}=f^\dagger_{s\sigma}f^{}_{s\sigma}$       \footnote{Considering
Gaussian   wave  packets  we   have:  $t=3\eta^2e^{-(\pi   \eta/2)^2}  E_{R}$,
$U=a_{s}k\sqrt{8/\pi}\eta^3$,  $U_{f}=U(V_{0}^{\prime}/V_{0})^{3/4}$,  $\Delta
\epsilon   =\left[   \eta^4(\gamma^4-1)-3\eta^2(\gamma^2-1)  \right]   E_{R}$,
$V=\sqrt{2}\beta  e^{-(\pi   \eta  \beta)^2/4}\;  t$,   with  $E_{R}=\hbar  ^2
k_{laser}^2/2m$,              $\gamma=\left(V_{0}^{\prime}/V_{0}\right)^{1/4}$,
$\beta=\gamma/(1+\gamma^2)^{1/2}$.}.

The Hamiltonian  (\ref{ham}) has  been extensively studied  in the  context of
strongly correlated  electrons. It is well  known that in the  regime in which
$U_{f}, \Delta\epsilon \gg t,V \gg U$, the so called Kondo regime \cite{Ueda},
strongly correlated effects appear.  Within this regime the low-energy physics
of Hamiltonian (\ref{ham}) can be described by an effective model in which the
$f$-atoms  degrees of  freedom are  represented by  localized spins,  the well
known Kondo lattice model (KLM)\cite{Fulde, Ueda}, \vspace{-0.35cm}
\begin{eqnarray}
H_{KLM}=-t\hspace{-0.1cm}
\sum_{<\ell,\ell^{\prime}>\sigma}c^{\dagger}_{\ell\sigma}c^{}_{\ell^{\prime}\sigma}
+ J \sum_{s}\bold{S}_{s}^{f}\cdot \bold{S}_{s}^{c},
\label{ham1}
\end{eqnarray}
where      $\bold{S}_{s}^{f}=\frac{1}{2}\sum_{\sigma^{},      \sigma^{\prime}}
\bold{\tau}_{\sigma^{},
\sigma^{\prime}}f^{\dagger}_{s\sigma^{}}f^{}_{s\sigma^{\prime}}$ are localized
spins             and            $\bold{S}_{s}^{c}=\frac{1}{2}\sum_{\sigma^{},
\sigma^{\prime}}\bold{\tau}_                                        {\sigma^{},
\sigma^{\prime}}d^{\dagger}_{s\sigma^{}}d^{}_{s\sigma^{\prime}}$           with
$d_{s\sigma}=\sum_{\langle  s,\ell \rangle}c_{\ell\sigma}$,  $\tau$  being the
vector of  Pauli matrices.  The exchange  interaction $J=2V^2/\Delta \epsilon$
is antiferromagnetic, and though typically very small ($J \ll t$) in condensed
matter systems, is  the source of interesting many-body  effects.  The KLM has
been studied typically  in two different situations \cite{Ueda}.   1) A single
localized spin weakly coupled to a continuum, which is the usual Kondo problem
\cite{Fulde, Hewson}.   Here it is well  known that for  temperatures bellow a
critical temperature  $T_{K}$, the Kondo  temperature, a many-body  singlet is
formed  composed of  the localized  spin and  the local  spin  polarization of
conduction electrons.  As  a result of singlet formation  a resonance appears,
the Kondo resonance,  which is responsible of many  interesting effects.  As a
remarkable  feature,   Kondo  effect  is  non-perturbative  in   $J$,  with  a
characteristic  exponential  behavior  of  the  Kondo  temperature  $T_{K}\sim
2te^{-1/2J\rho(\epsilon_{F})}$, where  $\rho(\epsilon_{F})$ is the  density of
states at the  Fermi level. The case  of a single impurity coupled  to a small
metallic grain has  been also theoretically studied in  \cite{Jan}. Here it is
found that the Kondo resonance can  be strongly affected by the finite size of
the  grain.  2) A  Kondo lattice  with typically  one impurity  per conduction
electron.   In  this  case  interaction  between  localized  spins  (the  RKKY
interaction, with  a characteristic long range oscillatory  behavior) can take
place, mediated by the continuum of conduction electrons.  Competition between
local Kondo singlet formation and  RKKY magnetism has been investigated as the
ratio $t/J$ is  varied \cite{Ueda}.  Since magnetism is  a perturbative effect
in $J$,  it is predicted  to dominate over  Kondo effect in the  weak coupling
regime ($J \ll t$) \cite{Ueda}.

The atomic  system that  we have under  consideration can realize  a situation
different from  the cases  described above.  Let  us assume that  tunneling is
only allowed along one direction  (potential barriers have been made very high
in the  other directions) and that  the parameters in  (\ref{ham}) fulfill the
conditions to  be in the  Kondo regime.  The  system realizes then a  1D Kondo
lattice in which impurities interact with each other through a discretized set
of levels.  A  new characteristic energy scale appears,  namely the separation
between levels  of an island,  $\Delta$.  For $\nu=1/2$ ($\nu=M/N$,  $N$ being
the number of particles and $M$  the total number of sites), the separation of
the Fermi  level and the  next excitation is $\Delta=2t\sin(\pi/L)$,  which is
finite in our  case.  In addition, the  ratio $t/J$ (as we show  later) can be
varied, so  that discussion  of both  the strong and  weak coupling  limits is
experimentally relevant in this case.

{\em Strong coupling  regime}: $J \gg \Delta$.  Kondo  screening dominates the
physics  of  the  problem.   Since  tunneling is  very  small,  impurities  are
basically  disconnected   from  each   other  and  singlet   formation  occurs
independently for each of them.  In  this limit we can use a generalization of
the variational wave  function of Varma and Yafet  \cite{Varma}
for the ground state:
\vspace{-0.25cm}
\begin{equation}
\hspace{-0.155cm}|\Psi \rangle=\prod_{s} \left( \beta + \sum_{k} \beta_{k} \left( f^{\dagger}_{s\uparrow}
             A^{}_{s k \uparrow}+
             f^{\dagger}_{s\downarrow}A^{}_{s k \downarrow}\right)
|FS\rangle_{s} \right),
\label{var}
\end{equation}
where   $A_{sk\sigma}=\sqrt{\frac{2}{L}}\sum_{\langle   s,\ell\sigma  \rangle}
\sin(k\ell)c_{\ell\sigma}$,   $k=\pi    n/L$,   $n=1,   \ldots,    L-1$,   and
$|FS\rangle_{s}=\prod_{k,\sigma}^{k_{F}}A^{\dagger}_{sk\sigma}$            with
$k_{F}=\pi  N/2L$.   The  variational  coefficients  $\beta$  and  $\beta_{k}$
satisfy $|\beta|^2+2\sum_{k}|\beta_{k}|^2=1$.   For each $f$-site, (\ref{var})
describes singlet formation  with a delocalized state of  momentum $k$ with an
amplitude  given  by  $\beta_{k}$.   Minimization  of  $E=\langle  \Psi|H|\Psi
\rangle/  \langle \Psi|\Psi  \rangle$ with  respect to  the  $\beta$'s yields:
$\beta_{k}/\beta=-\frac{2}{\sqrt{L}}V\sin  k/(T_{K}+\Delta_{k})$, and  a Kondo
temperature given by the implicit equation
\begin{equation}
T_{K}=\Delta  \epsilon - \epsilon(k_{F})  +  
\frac{16V^2}{L}  \;\sum_{k}^{k_{F}}\frac{\sin
^2k} {T_{K} + \Delta_{k}},
\label{Tkondo}
\end{equation}
where   $\Delta_{k}=\epsilon(k_{F})-\epsilon(k)$,   $\epsilon(k)=-2t\cos   k$.
Under realistic conditions  we will have $L<10$ and  the sum in (\ref{Tkondo})
will be a  discrete sum with a few  ($\sim L/2 $ for $ \nu=1/2$)  terms.  As a
result, $T_{K}$ does  not go to zero with $t/J$ (Figure  2), but remains $\sim
J$ due to the finite size  of the conducting island.  When $\Delta$ becomes of
the order of $J$ the size of the singlets becomes comparable to the separation
$L$ between  supersites, so  that the screening  cloud of one  impurity starts
affecting the next supersite. An  interplay between Kondo effect and magnetism
takes place.
\begin{figure}[h]
\includegraphics[height=5.95cm,width=6.5cm]{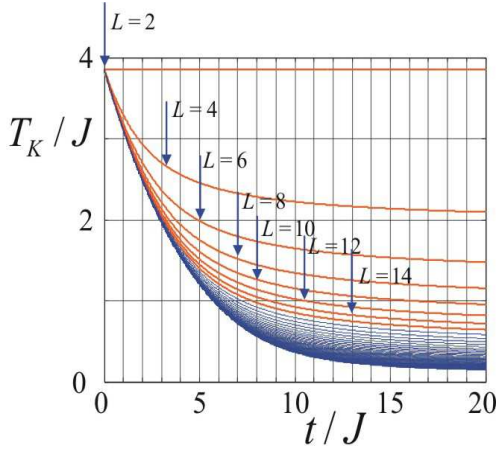}
\caption{Kondo temperature $T_{K}$ as a function of $J/t$. Different curves
correspond  to increasing  values of  $L$ ($N=L/2$).  For realistic  cases (in
orange),  the arrows mark  points at  $T_{K}=2\Delta$, to  the right  of which
expression (\ref{Tkondo}) is not valid.}
\end{figure} 

{\em  Weak coupling  regime} $\Delta  \gg  J$.  A  very different  situation
corresponds to  the regime in which  the spacing between energy  levels in the
conducting islands  ($\sim 2t\sin(\pi/L)$) becomes  much larger that  the Kondo
temperature ($\sim J$).   Within this limit atomic orbital  degrees of freedom
are completely frozen,  with excitations above the Fermi level  in each of the
islands  taking  part of  the  problem  only  as virtual  states.   Performing
adiabatic elimination  of these  excitations in Hamiltonian  (\ref{ham1}), we
obtain  an effective  Hamiltonian  for the  spin  degrees of  freedom.  As  an
interesting feature  the resulting  Hamiltonian depends on  the parity  of the
number of particles per conducting island, $N_{c}$.  This even-odd effect is a
clear manifestation of the finite size of the conducting islands.

{\em $N_{c}$=even}. The  Fermi level of each island is  occupied by two atoms.
In this case the only spin degrees of freedom correspond to atoms localized in
supersites.   An  effective   spin-spin  interaction  between  neighboring
supersites appears, mediated by the Fermi sea in between them. To second order
perturbation theory in $J$, \vspace{-0.1cm}
\begin{equation}
H_{SS}=J_{eff}^{M} \sum_{\langle s,s^{\prime}\rangle}
\bold{S}_{s}^{f}\cdot \bold{S}_{s^{\prime}}^{f},
\label{ham2}
\end{equation}
where          $J_{eff}^{M}=\frac{J^2}{tL}         \sin(k_{F})\sin(k_{F}(L-1))
\sin(k_{F}+\Delta k)  \sin((k_{F}+\Delta k)(L-1))$, and  $\Delta k=\pi/L$.  We
see that Kondo effect disappears and magnetism is induced for localized atoms,
the ground  state being antiferromagnetic  or ferromagnetic depending  on both
$L$ and $k_{F}$.  We note that due to the characteristic topology imprinted by
the superlattice, Heisenberg (and not RKKY) magnetism is induced.

{\em $N_{c}$=odd}.   The Fermi level of  each island is occupied  by one atom,
whose  spin  comes into  play.   The effective  Hamiltonian  is  in this  case
$H=J_{eff}^{K}  \sum_{s}  \bold{S}_{s}^{f}\cdot  \bold{S}_{s}^{k_{F}}$,  where
$\bold{S}_{s}^{k_{F}}=\frac{1}{2}\sum_{\sigma^{}, \sigma^{\prime}}\bold{\tau}_
{\sigma^{},
\sigma^{\prime}}A^{\dagger}_{sk_{F}\sigma}A^{}_{sk_{F}\sigma^{\prime}}$,    and
$J_{eff}^{K}=4J/L$.  Kondo  effect remains in  this case and magnetism  is not
induced. The  ground state consists  in this case  of singlets formed  by each
localized atom  and the atoms at  the Fermi level in  neighboring islands. The
Kondo temperature is $T=2J_{eff}^{K}$.

{\em  Numerical  results}.   To  illustrate  the  predictions  above  we  have
numerically diagonalized Hamiltonian (\ref{ham})  for a small 1D superlattice.
In Fig.   3 and Fig.  4  we plot the spin-spin  correlation functions $\langle
\bold{S}_{f}\cdot  \bold{S}_{\ell} \rangle$  (spatial correlation  of  a fixed
$f$-spin  with  the  rest  of  sites  in  the  chain,  $\ell$),  and  $\langle
\bold{S}_{f}\cdot \bold{S}_{k} \rangle$ (correlation  of a fixed $f$-spin with
a delocalized spin with momentum $k$) for the exact ground state.  We consider
different cases. a)  $L=4$, $N_{c}=4$ (Fig. 3). Fig. 3a)  shows a clear smooth
transition from local Kondo singlet formation to magnetism of localized spins.
For  small values of  $t/J$ each  localized $f$-spin  is antiferromagnetically
correlated with its next neighboring  sites (forming a singlet with them).  As
$t/J$  increases correlations of  each impurity  with its  neighboring islands
disappear,  at the  same time  that correlations  between next  supersites are
induced.   The transition  (arrow in  inset of  Fig. 3a))  takes  place around
$T_{K}\sim 2\Delta$  ($t/J\sim3.25$) as  predicted.  As stated  by Hamiltonian
(\ref{ham2})      impurities       are      antiferromagnetically      coupled
($J_{eff}^{M}=-J^2/16t$).  b)$L=4$, $N_{c}=3$ (Fig.  4).  Kondo effect appears
in this case as $t/J$ increases. In real space Fig. 4a) shows how singlet-type
correlations become more  and more extended along the  conducting islands next
to  each   impurity,  whereas  neighboring   impurities  remain  uncorrelated.
Delocalization of  the singlet  becomes more evident  in momentum  space (Fig.
4b), where a resonance, the Kondo  resonance, appears at the Fermi level.  The
Kondo  temperature is  always of  the order  of $J$  (see inset  of  Fig. 4b),
reaching the  limiting value  $T\sim 2J_{eff}^{K}=2J$ for  $J \ll  \Delta$, as
predicted above.
\begin{figure}[t]
{\Large a)}
\includegraphics[height=4.78cm,width=7cm]{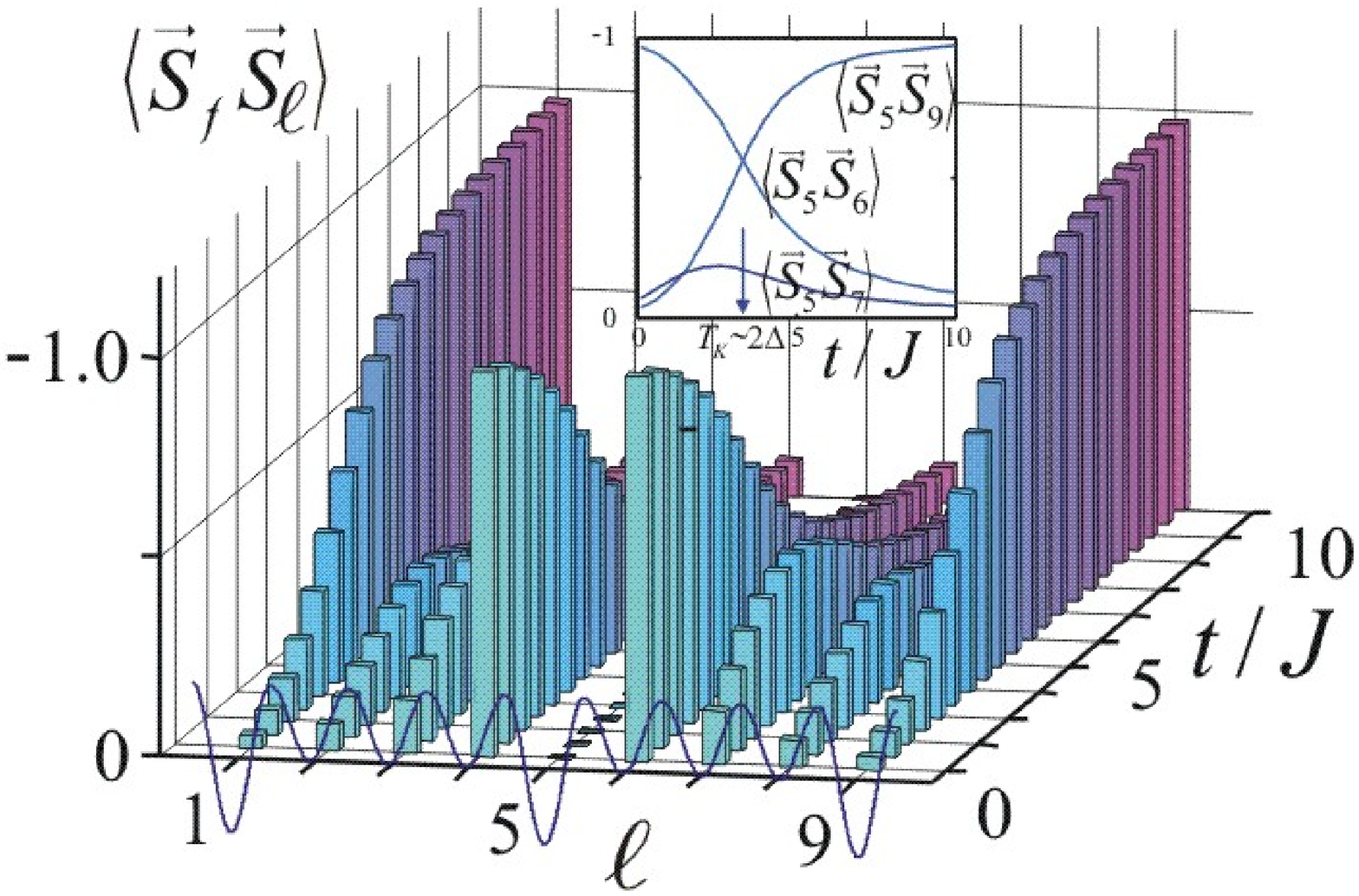}\\
\vspace{0.5cm}
{\Large b)}
\includegraphics[height=4.78cm,width=7cm]{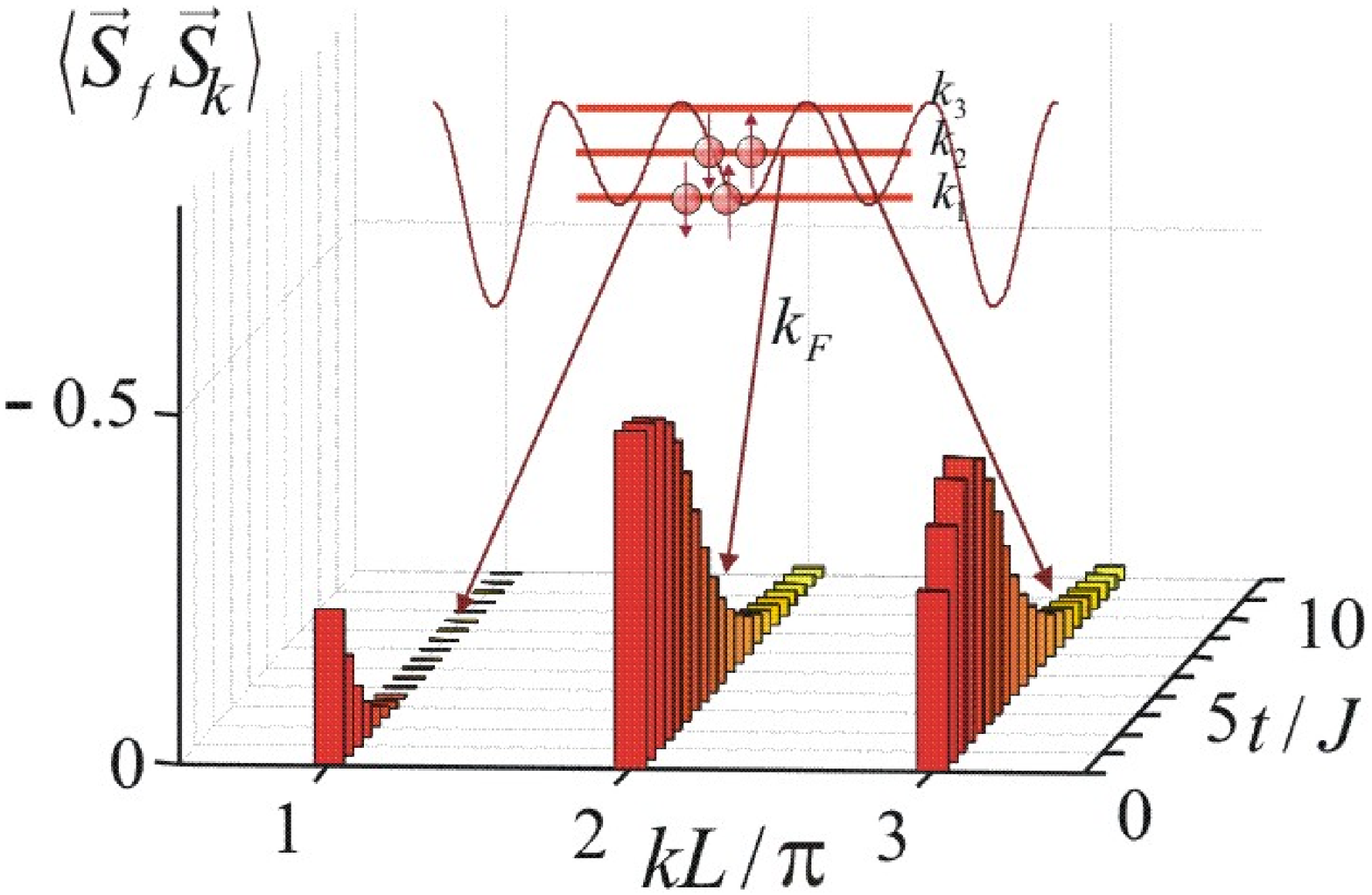}
\caption{Spin-spin correlation functions 
of a fixed f-site
($s=5$) with the rest of sites (a) and the k-momentum states of
a neighboring island (b), as a function of $t/J$. Parameters:
$M=9$, $N=11$ ($N_{c}=4$), $U=0$, $U_{f}=10 \Delta \epsilon$, 
$\Delta \epsilon=10J$. In the inset of a) correlations of supersite
$5$ with sites $6$, $7$, and supersite $9$ are plotted.}
\end{figure}
\begin{figure}[t]
{\Large a)}
\includegraphics[height=4.78cm,width=7cm]{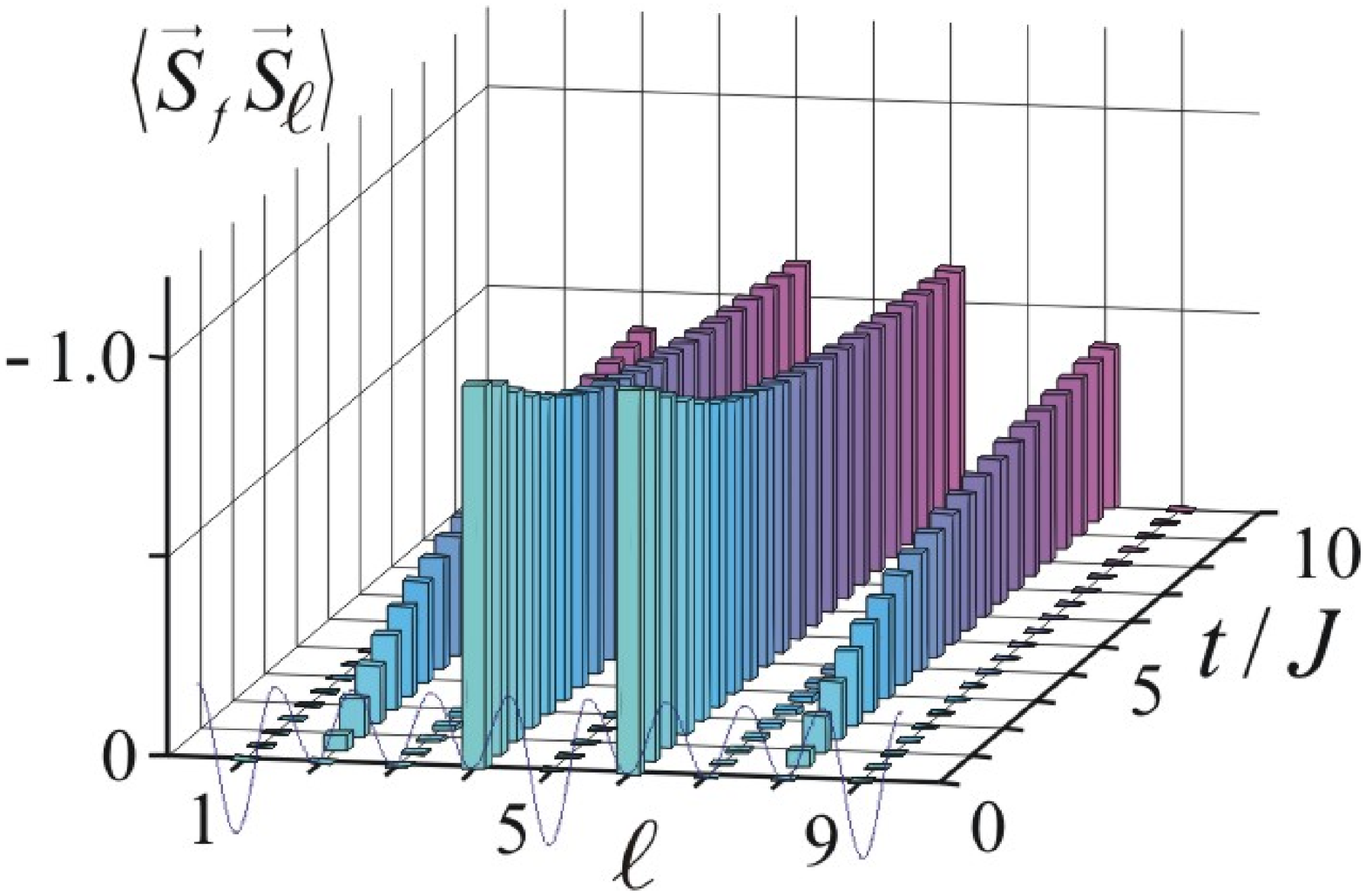}\\
\vspace{0.5cm}
{\Large b)}
\includegraphics[height=4.78cm,width=7cm]{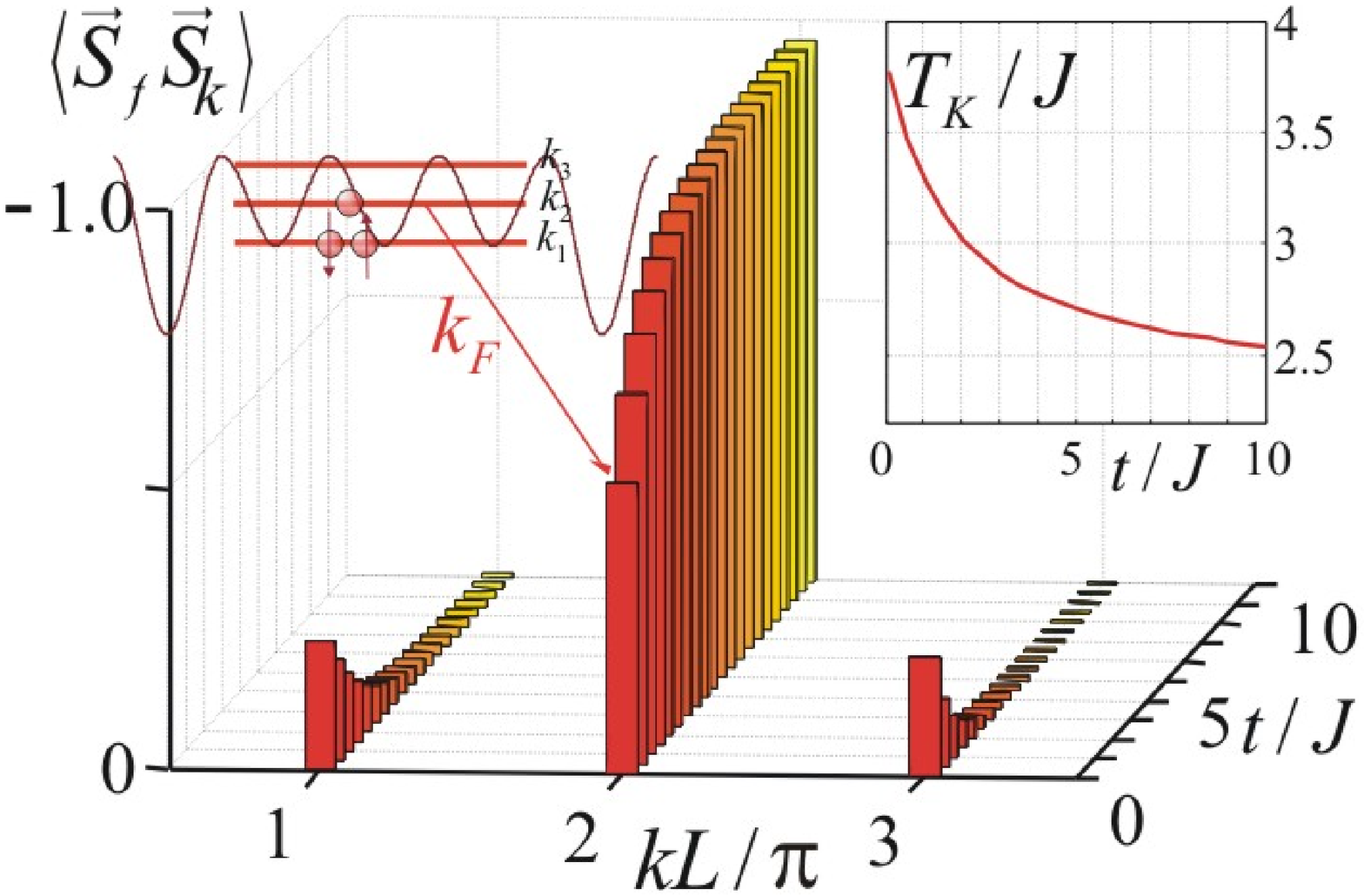}
\caption{Same as in Fig. 3 with $N=9$ ($N_{c}=3$).
The Kondo temperature is plotted in the inset of b).}
\end{figure} 

We discuss  now the  experimental realization of  the regimes we  have studied
above.   First   of  all  we  need   $U_{f}  \gg  U$.    Since  $U_{f}/U  \sim
V_{0}^{\prime}/V_{0}$ the  potential depth of  supersites must be  very large.
As a consequence the energy offset $\Delta \epsilon \sim V_{0}^{\prime}-V_{0}$
will be also large whereas the coupling $V$ will be very small.  For a typical
value of  $V_{0}^{\prime}/V_{0} \sim 10$ we obtain  $J<10^{-4}t$, which yields
extremely  small  Kondo  temperatures,  completely  unrealistic  with  current
technology.   In order  to  overcome  this problem  we  propose the  following
scheme.   Let  us consider  an  ensemble of  $^{6}$Li  atoms,  which have  six
hyperfine  states $|F,M_{F}  \rangle$, with  a  total spin  $F=1/2$, or  $3/2$
\cite{Li}. We  consider a  situation in which  four of these  internal states,
$a\uparrow, a\downarrow, b\uparrow, b\downarrow $ are trapped (as shown in Fig
5.).  We will also assume that interactions are engineered in such a way
that interactions  of type $a-a$ and $a-b$  are negligible, whereas $b-b$
are  positive and  large.  The  goal is  to use  a-atoms  (non-interacting) as
``conducting''  c-atoms,  and  b-atoms  (strongly  interacting)  as  localized
f-atoms. One  possibility is  the following.  Let  us assume we  load $^{6}$Li
atoms in an  optical lattice, with $N_{\uparrow}=N_{\downarrow}$, $\nu_{b}=1$,
and  $\nu_{a}<1$ (variable).   Atoms of  type  $a$ will  delocalize along  the
lattice  forming a  Fermi sea,  whereas atoms  of type  $b$ will  be localized
forming  a Mott phase.   If the  superlattice is  now adiabatically  turned on
supersites will  be filled  up, each supersite  containing two  $a$-atoms with
opposite  spins and  only one  $b$-atom. The  idea is  to couple  $a$-atoms in
normal sites with $b$-atoms in  supersites by using an off-resonant laser.  By
tuning the intensity  $\Omega$ and frequency $D$ of  the laser, the parameters
$\Delta  \epsilon$ and  $V$ can  be  tuned to  the desired  values.  The  only
restriction here  is that the  laser must not  excite other processes,  as for
example, transitions between  a-atoms and b-atoms in normal  sites.  This sets
the condition $\left(  \Omega/\delta \right)^{2} \ll 1$ (see  Fig. 5).  Taking
this into account it is possible  to make $0.5<t/J<20$, with values of $J \sim
0.1 E_{R}$.  This  yields Kondo temperatures $T_{K} \sim \mu  K$, which are of
the order of the ones required to observe superfluidity for fermionic atoms in
optical  lattices  \cite{Lukin}.   Finally,  we  discuss  realization  of  the
superlattice.    Each  of   the   Fourier  components   of  the   superlattice
($e^{ik_{n}x}$,   with   $k_{n}=k  n/L$,   $n$   integer)   can  be   realized
experimentally  by two counter-propagating  laser beams  with wave  vector $k$
forming  an appropriate angle  $\theta_{n}$.  Therefore  the number  of lasers
required for  a good realization of  the superlattice increases  with $L$ (for
$V_{0}^{\prime}/V_{0}\sim 4$ a  set of $\sim 2L$ lasers is  needed) But, as we
have  shown, systems  with  $L=3,4$ already  display  the strongly  correlated
phenomena we have predicted.
\begin{figure}[h]
\includegraphics[height=4cm,width=6.4cm]{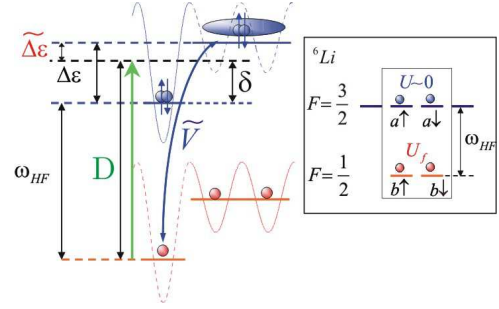}
\caption{Experimental scheme (see text). The renormalized
parameters are $\widetilde{\Delta \epsilon}=
\omega_{HF}+\Delta \epsilon-D$ (with $\omega_{HF}$ the hyperfine frequency), 
and $\widetilde{V}=\Omega V$.}
\end{figure}

{\em Experimental observation}. One of  the most clear manifestations of Kondo
effect  is the  appearance  of Kondo-induced  resonances  in several  physical
magnitudes,   as  for   instance  spin-spin   correlation   functions.   These
correlations may be however hard  to measure in actual experiments, since they
involve  two  particle  correlations.   Instead,  we propose  to  measure  the
one-particle             correlation             function             $\langle
A_{k\sigma}^{\dagger}f_{\sigma}\rangle$.   From  (\ref{var})  we get  $\langle
A_{k\sigma}^{\dagger}f_{\sigma}\rangle  \sim  \beta  \beta_{k}$, so  that  the
localization  of  the  singlet in  the  vicinity  of  the Fermi  level  (Kondo
resonance) will show up in this quantity.  This correlation can be obtained in
the  following   way.   Both  $\langle   c_{\sigma  \ell}^{\dagger}  c_{\sigma
\ell^{\prime}}^{}\rangle$    and  $\langle
f_{\sigma s}^{\dagger} f_{\sigma s^{\prime}}^{}\rangle$ can be detected in the
interference  pattern measured  after  free expansion  of  $a$ and  $b$-atoms,
respectively.  As  well, by applying a  $\pi/4$ laser pulse 
between the  internal states $a$
and    $b$  right before measurement,    $\langle
\left(f^{\dagger}+c^{\dagger}\right) (f+c)\rangle$ can be obtained.  Combining
these  three we obtain  the desired  correlation.  Finally,  magnetism between
impurities can  be detected  by using spin  dependent Bragg  scattering, which
will show  up the  antiferromagnetic or ferromagnetic  order of  the localized
spins.

In  conclusion we  have shown  that fermionic  atoms in  optical superlattices
exhibit strongly correlated phases,  from Kondo singlet formation to magnetism
of localized spins. Characteristic features of this system are the finite size
of the conducting islands coupled  to supersites, which strongly influence the
competition between  Kondo effect and magnetism. These  entangled phases could
be used for atomic spintronics and spin-filtering.

Discussions with G. G\'omez-Santos and J. von Delft are gratefully
acknowledged.

\end{document}